\documentclass[article]{jss}
\usepackage{amsmath}
\usepackage{algpseudocode}
\usepackage{algorithm}

\author{Florence Bockting \\TU Dortmund University \And 
         Paul-Christian Bürkner\\TU Dortmund University}
\title{\pkg{elicito}: A Python Package for Expert Prior Elicitation}

\Plainauthor{Florence Bockting, Paul-Christian Bürkner} 
\Plaintitle{elicito: A Python Package for Expert Prior Elicitation} 
\Shorttitle{\pkg{elicito}: A Python Package for Expert Prior Elicitation} 

\Abstract{
  Expert prior elicitation plays a critical role in Bayesian analysis by enabling the specification of prior distributions that reflect domain knowledge. However, expert knowledge often refers to observable quantities rather than directly to model parameters, posing a challenge for translating this information into usable priors. We present \pkg{elicito}, a Python package that implements a modular, simulation-based framework for expert prior elicitation. The framework supports both structural and predictive elicitation methods and allows for flexible customization of key components, including the generative model, the form of expert input, prior assumptions (parametric or nonparametric), and loss functions. By structuring the elicitation process into configurable modules, \pkg{elicito} offers transparency, reproducibility, and comparability across elicitation methods. We describe the methodological foundations of the package, its software architecture, and demonstrate its functionality through a case study.
}
\Plainkeywords{keywords, comma-separated, not capitalized, Java} 

\Address{
  Florence Bockting\\
  Department of Statistics\\
  Chair of Computational Statistics\\
  TU Dortmund University\\
  44227 Dortmund, Germany\\
  E-mail: \email{florence.bockting@tu-dortmund.de}\\
  URL: \url{https://florence-bockting.github.io/}
}

\begin{document}

\section[Introduction]{Introduction}

The goal of \emph{expert prior elicitation} is to specify prior distributions for model parameters or events \citep{falconer2024eliciting,mikkola2024preferential} in Bayesian models that accurately reflect the expectations of a domain expert \citep[for a recent review, see][]{mikkola2023prior}. A central difficulty of this objective is that expert knowledge often relates indirectly, or in a non-transparent way, to the model quantities for which priors are required. To address this challenge, the research field \emph{prior elicitation} has emerged, originating in the 1960s \citep{winkler1967assessment, kadane1980interactive, kadane1998experiences}, and continues to evolve as an active area of study \citep{stefan2022practical, mikkola2023prior, falconer2022methods}.

Two major strands can be distinguished within this field, depending on whether the focus is on the extraction or the translation process of expert knowledge \citep{manderson2023translating}. The \emph{extraction process} focuses on the interaction with the expert, including identifying which types of questions can reasonably be posed, understanding cognitive heuristics that may influence expert judgments, training experts to effectively quantify their knowledge, and recommending a set of quantities that can be meaningfully elicited \citep{dias2018elicitation, stefan2022practical, european2014guidance}. In contrast, the \emph{translation process} focuses on understanding the statistical and mathematical relationships between the model parameters, for which a prior distribution is required, and any other model-derived quantities about which the expert might provide information, such as the outcome variable or summary statistics derived from them \citep{hartmann2020flexible, da2019prior, manderson2023translating, perepolkin2023quantile}.

Insights from both domains have informed the development of prior \emph{elicitation methods} \citep{mikkola2023prior}. These methods can be broadly classified into structural and predictive elicitation approaches \citep{kadane1998experiences, falconer2022methods}. In \emph{structural} elicitation methods, expert knowledge is directly expressed in terms of the parameters of a Bayesian model. As a result, these methods primarily focus on the fitting task, namely, fitting a prior distribution given the expert data \citep{kadane1980interactive, bedrick1996new}. In contrast, \emph{predictive} elicitation methods elicit expert beliefs about the outcome variable rather than the model parameters. In this case, the methods must address not only the fitting task but also a translation task: converting beliefs about observable outcomes into corresponding priors over model parameters \citep{akbarov2009probability, muandet2017kernel, mikkola2023prior, hartmann2020flexible, manderson2023translating,da2019prior,perepolkin2024hybrid}.

The entire \emph{elicitation process}, which encompasses the workflow from extracting knowledge from a domain expert to learning prior distributions for the model quantities of interest, can be summarized in four stages \citep{garthwaiteStatisticalMethodsEliciting2005}: The \emph{setup stage} involves preparing for the main elicitation, including the identification of suitable expert(s) and the selection of relevant information to be elicited. This is followed by the \emph{elicitation stage}, during which the selected information is extracted from the expert(s) using structured elicitation protocols. In the subsequent \emph{fitting stage}, the elicited information is used to derive corresponding prior distribution(s). The final \emph{evaluation stage}, entails assessing the plausibility of the learned priors in collaboration with the expert(s). If discrepancies or inconsistencies are identified, this stage may prompt revisions of decisions made in earlier stages and repetition of the process to obtain more accurate or reliable priors.

In our previous work, we focused on the process of translating expert knowledge into prior distributions. Specifically, building on the work of \citet{hartmann2020flexible, manderson2023translating, da2019prior}, we introduced a simulation-based approach for learning prior distributions of model parameters from different types of expert input ranging from the parameter to the observable space \citep{bockting2023simulation}. In a recent extension, we increased the flexibility of our simulation-based approach by also supporting different types of prior distributions from joint non-parametric to independent, parametric priors \citep{bockting2024expert}.

Building on this work, we developed a modular framework that incorporates all core components of a prior elicitation method. Each module in the framework can be customized which allows for the support of a wide range of elicitation methods that may differ in their choice of generative models, the type and structure of expert information, the nature of prior assumptions (including both parametric and nonparametric priors), and the specification of loss functions. We translated this conceptual idea into an algorithmic design implemented via the Python package \pkg{elicito} which is introduced in this paper. We think that the possibility to develop different elicitation methods within the same framework offers several advantages. It increases transparency regarding the decisions required to implement a given elicitation method, facilitates the comparison of different methods, and enables the use of common diagnostics to evaluate both method performance and the resulting prior distributions.

The remainder of this paper is organized as follows:
Section~\ref{sec:method} provides an overview of the methodological and statistical foundations underpinning our software implementation, \pkg{elicito}. Section~\ref{sec:software} presents the software architecture, detailing each module in relation to the conceptual stages of the prior elicitation process. In Section~\ref{sec:case-study}, we demonstrate the application of \pkg{elicito} through a simple case study that illustrates its functionality in a concrete context. Section~\ref{sec:relevant-software} reviews related software packages currently available in the field and discusses their relevance to the proposed implementation. Finally, Section~\ref{sec:discussion-and-conclusion} concludes the paper with a general discussion and an outline of future developments of \pkg{elicito}.

\section[Method]{Method}
\label{sec:method}
In the following section, we present a conceptual overview of our simulation-based prior elicitation framework, as introduced in \citet{bockting2023simulation} and \citet{bockting2024expert}. Readers already familiar with this workflow may wish to skip directly to Section~\ref{sec:software}, which focuses on the implementation details of the \pkg{elicito} software.

The overall workflow of our framework follows the five stages of the prior elicitation process as outlined by \citet{garthwaiteStatisticalMethodsEliciting2005}, and is operationalized through the following six steps:
\begin{enumerate}
    \item \emph{Define the generative (statistical) model}: Specify the generative model, including dimensionality and parameterization of the prior distribution(s). (Setup stage, Section~\ref{subsec:generative-model})
    \item \emph{Identify the information to be elicited and the corresponding elicitation techniques}: Determine the relevant expert knowledge to be elicited and select appropriate elicitation techniques. (Setup stage, Section~\ref{subsec:target-quantities-and-elicited-statistics})
    \item \emph{Extract information from the expert and simulate corresponding predictions from the generative model}:
    Collect expert input and simulate predictions from the generative model. Perform all necessary computational steps to generate model predictions corresponding to the expert-elicited information. (Elicitation stage, Section~\ref{subsec:expert-elicited-and-model-implied-statistics})
    \item \emph{Evaluate consistency between expert information and model predictions}: Quantify the discrepancy between expert-provided information and model predictions using a multi-objective loss function. (Fitting stage, Section~\ref{subsec:prior-hyperparameter-optimizsation})
    \item \emph{Learn prior distributions consistent with expert information}: Use an optimization algorithm to minimize the loss function, thereby identifying prior distributions that align with the elicited expert information. (Fitting stage, Section~\ref{subsec:prior-hyperparameter-optimizsation})
    \item \emph{Evaluate the learned prior distributions}: Repeat the optimization with different random seeds to assess the variability of priors that fit the expert data. Select a plausible prior distribution in collaboration with the domain expert or apply model averaging techniques.  (Evaluation stage, Section~\ref{subsec:evaluation-learned-prior})
\end{enumerate}
The following sections provide a detailed explanation of each step. An overview of the symbols used throughout the text, along with their descriptions, is provided in Appendix~\ref{app:symbols}.

\subsection{Generative Model and Prior Assumptions}
\label{subsec:generative-model}
The foundation of our framework is the generative model $\mathcal M$, which defines a statistical model over the joint probability distribution $p(y,\theta)$, where $y$ denotes the observed data and $\theta$ represents the unobserved parameters \citep{haines2023from}. Within the Bayesian setting considered here, specifying the generative model further requires assumptions regarding the dimensionality, dependence structure, and parametric form of the prior distribution $p_\lambda(\theta)$ over the model parameters, where $\lambda$ denotes the hyperparameters of the prior. 

The interpretation of $\lambda$ depends on the nature of the prior. Under a  \emph{parametric prior}, $\lambda$ corresponds to the hyperparameters of the chosen parametric distribution family. In contrast, when a \emph{non-parametric prior} is employed, our framework utilizes deep generative models to learn the functional form of the prior distribution. In this case, $\lambda$ denotes the parameters of the deep generative model, typically the weights of a (deep) neural network.

The generative model $\mathcal M$ is specified by the user with the aim of learning prior distributions over the model parameters. Since the priors are parameterized by the hyperparameter vector $\lambda$, the objective of the elicitation framework is to identify a suitable value for $\lambda$. 
The guiding criterion for this learning task can be formualted as follows: When the generative model $\mathcal M \mid \lambda$, conditioned on a specific hyperparameter vector, is executed in forward mode, the resulting simulated model-derived quantities should align with the expectations specified by a domain expert, which are detailed in the following sections.

\subsection{Target Quantities and Elicited Statistics}
\label{subsec:target-quantities-and-elicited-statistics}
To enable the learning task, it is necessary to define a set of expert-informed criteria, or in other words, the expectations of the domain expert quantified through specific summary statistics. Ideally, this set should be both, \emph{interpretable} to domain experts and \emph{informative} for the generative model \citep{mikkola2023prior}. 
In practice, achieving both objectives simultaneously is challenging, as they are often not aligned. For example, information directly referring to model parameters can be highly informative for the learning process, yet frequently lacks interpretability from the perspective of a domain expert \citep{kadane1980interactive}.
By contrast, information related to the outcome variable is typically highly interpretable for domain experts but offers limited utility in constraining the individual hyperparameters $\lambda$, often resulting in poorly informed or weakly constrained prior distributions \citep{stefan2022practical}.
The suitability of a particular set of expert information is highly dependent on both the structure of the generative model and the specific expertise of the individual being queried \citep{denham2007geographically}. Consequently, an effective elicitation method should be capable of accommodating a wide range of different forms of expert knowledge.

Within our framework, expert information is specified through a two-stage process. First, we define \emph{what} type of information is to be elicited from the expert. This information is formalized as a function of the hyperparameters, denoted by $z_i=c_i(\lambda)$ for $i=1,\ldots,I$, and referred to as \emph{target quantities}. Typically, multiple target quantities ($I$ in total) are elicited from the user to inform different aspects of the data-generating process.
For instance, if we want to query the expert regarding the outcome variable (i.e., the observable space), one of the target quantities can be simply defined as $z_i=y$. This is obtained by defining the function $c_i$ as a query to the prior predictive distribution $p(y) = \int p(y \mid \theta) p_\lambda(\theta) d \theta$, where the likelihood is marginalized over the prior. Note that this expression is not just a constant as it refers to the prior predictive distribution and should not be confused with the marginal likelihood of observed data (see for this point also \citet{hartmann2020flexible}).
As another example, expert knowledge may be elicited about the parameter space. In this case, the target quantity is defined as $z_i=\theta_k$, which is obtained by defining $c_i$ as a simple projection onto $\theta_k$. 
In this way, we can also accommodate any other target quantity that can be derived from the model parameters or the data \citep[e.g., $R^2$; the proportion of variance explained by the model;][]{gelman2019r}.

Second, we define \emph{how} each target quantity is to be queried by selecting an appropriate elicitation technique. In our elicitation method, we represent an elicitation technique as a function $f_j$ for $j=1,\ldots,J$, of a target quantity $z_i$. The resulting set of \emph{elicited} target quantities is referred to as \emph{elicited statistics}, ~$\{t_m(\lambda)\}$, where $t_m(\lambda) = f_j(z_i)$ and $m$ refers to the corresponding $i \times j$ combination. In the following, we sometimes omit the explicit dependence on $\lambda$ and simply write $t_m$ instead of $t_m(\lambda)$. Note also that, depending on the elicitation technique employed, $t_m$ may accept either a single value or a set of values as input. 
To illustrate this point more clearly, consider the case in which we elicit expert knowledge about the predictive distribution conditional on a category of a categorical predictor, i.e., $z = y\mid \text{cat}$. To elicit this information, we query the expert regarding the $25\%, 50\%$, and $75\%$ quantiles of this predictive distribution. Thus, we have $t = \{Q_{25\%}(y\mid \text{cat}), \, Q_{50\%}(y\mid \text{cat}), \, Q_{75\%}(y\mid \text{cat})\}$.

\subsection{Expert-elicited and model-implied statistics}
\label{subsec:expert-elicited-and-model-implied-statistics}
Once the target quantities and elicitation techniques have been specified, the corresponding quantities of interest must be obtained from the domain expert through an appropriate interrogation. This yields a set of expert-elicited summary statistics, denoted by  $\{\hat t_m\}_{m=1}^M$. 
Similarly, a corresponding set of model-implied statistics, denoted by $\{t_m\}_{m=1}^M$, can be generated by simulating the generative model in forward mode and applying the introduced transformation $t_m = f_j(c_i(\lambda))$.

Given the expert-elicited statistics $\{\hat{t}_m\}$ and the model-implied statistics $\{t_m\}$, their agreement can be assessed using a discrepancy measure $\mathcal{D}_m(t_m(\lambda), \hat{t}_m)$ for $m=1,\ldots,M$. The choice of discrepancy measure may vary across statistics, depending on the nature and scale of each elicited statistic. 
Each weighted discrepancy term is referred to as a \emph{loss component} $L_m$, where $w_m$ represents the corresponding weight. The total loss is defined as the sum over all individual loss components, resulting in a \emph{multi-objective loss function} that quantifies the overall mismatch between expert knowledge and model simulations. Formally, this is given by
\begin{equation} 
    \mathcal L(\lambda) = \sum_{m=1}^M  \underbrace{w_m\cdot \mathcal D_m\left(t_m(\lambda), \hat t_m\right)}_{L_m}.
\end{equation}
The learning task of identifying a hyperparameter vector $\lambda$ such that the model-derived quantities align with expert expectations can thus be reformulated as an optimization problem
\begin{equation}\label{eq: loss}
    \lambda^*= \min_\lambda \mathcal L(\lambda)
\end{equation}
which seeks the hyperparameter values $\lambda^*$ that minimize the multi-objective loss function. 

\subsection{Prior hyperparameter optimization}
\label{subsec:prior-hyperparameter-optimizsation}
To minimize the total loss $\mathcal{L}(\lambda)$, we formulate the problem as an optimization task and solve it using mini-batch stochastic gradient descent (SGD), as defined in Equation~\ref{eq: loss}. This approach requires computing the gradient of the loss function with respect to the hyperparameter vector~$\lambda$, followed by iterative updates of $\lambda$ in the direction opposite to the gradient \citep{goodfellow2016deep}.
Gradient computation is enabled via automatic differentiation using the explicit or implicit reparameterization trick \citep{kingma2014auto,figurnov2018implicit}.

The updated hyperparameters $\lambda'$ are then used to generate new samples from the prior distribution $p_{\lambda'}(\theta)$, from which an updated set of model-implied statistics $\{t_m(\lambda')\}$ is computed. 
The discrepancy between these updated model-implied statistics and the expert-elicited statistics is then reassessed. Based on the resulting loss, the hyperparameters are further adjusted using the mini-batch SGD procedure. This iterative process continues until a predefined convergence criterion is met or a specified stopping condition is reached.

\subsection{Evaluation of learned prior distribution}
\label{subsec:evaluation-learned-prior}
As an initial step, the learning process is assessed for unexpected behavior that may indicate issues during optimization. If no such irregularities are observed, the learned prior distribution can be evaluated.

\paragraph{Monitoring the learning process}
A first assessment of the learning progress involves examining the trajectory of the total loss, along with the individual loss components, throughout the optimization process. Ideally, the total loss should exhibit a generally decreasing trend, often resembling exponential decay, and eventually stabilize near zero \citep{goodfellow2016deep}.
However, it is important to note that the \emph{individual} loss components may temporarily increase before decreasing, particularly in the presence of conflicting objectives within the multi-objective loss function. The shape and slope of the loss trajectory can offer insights into the effectiveness of the optimization procedure. For example, an excessively steep slope may indicate an overly large learning rate, which can cause the optimizer to overshoot local or global minima. Conversely, a learning rate that is too small may result in negligible parameter updates per iteration, leading to slow convergence and prolonged training times \citep{buduma2017fundamentals}.

In addition to monitoring the loss, the update trajectory of the hyperparameters $\lambda$ can offer valuable insights into the optimization process. Ideally, the trajectory should stabilize over successive iterations, converging to a specific value. However, in some cases, inspecting individual hyperparameters is impractical due to their high dimensionality. This is particularly true for non-parametric priors, where $\lambda$ correspond to the weights of deep neural networks.
In such settings, it more feasible to monitor the evolution of specific summary characteristics of the prior distribution across optimization iterations. These may include location (e.g., mean), scale (e.g., variance), skewness, or correlations between marginal priors. Visualizing the trajectories of these components can provide insight into the stability and convergence of the learning process.  

Regardless of whether individual hyperparameters or summary characteristics of the prior are examined, a failure of the trajectory to converge or persistent high variability may signal problems in the learning process. Such behavior can stem from an inappropriate learning rate or from insufficient information in the elicited statistics to meaningfully constrain the prior. For example, a collapse of the prior variance toward zero during training may indicate that the available expert information is inadequate to guide the learning process, resulting in overly narrow or degenerate prior distributions.

Another useful diagnostic for assessing the success of the learning process is to directly compare the expert-elicited statistics with the final model-implied statistics. While the individual loss components already quantify the discrepancies between these two sets of quantities, this additional diagnostic offers a complementary perspective by evaluating the match in the original scale of the elicited statistics. This can make it easier to identify meaningful mismatches. In contrast, interpreting the loss trajectory alone can be challenging, as it is not always evident whether a near-zero loss is small enough in practical terms. 

\paragraph{Inspecting the learned prior}
Once the inspection of the optimization process is complete and no unexpected behavior is observed, the learned prior distributions can be assessed in detail. This assessment may include visual inspection of the marginal prior distributions, as well as examination of the joint prior through the analysis of the correlation structure between marginal components. 

An important consideration is that the elicited statistics used to guide the learning of the prior are typically insufficient to fully identify the generative model. As a result, there is not a unique hyperparameter vector $\lambda^*$ that matches the set of elicited statistics $\{\hat{t}_m\}$; instead, there exists a set of equally valid solutions, $\{\lambda^*\}_{k=1}^K$. Consequently, this gives rise to a corresponding set of prior distributions $\{p_{\lambda^*}(\theta)\}_{k=1}^K$, all of which are equally consistent with the elicited expert knowledge.

To understand the variability among supported prior distributions, it is instructive to sample from the set of learned priors. This can be achieved by running the optimization process multiple times, each with different initializations, and subsequently inspecting the range of resulting priors. Together with the domain expert, these priors can be evaluated for plausibility. If some of the learned priors are deemed implausible, this suggests that the current elicited information is insufficient to adequately constrain the solution space, and additional constraints or elicited statistics should be incorporated into the loss function. If, on the other hand, the resulting set consists of multiple priors that are all considered plausible, one may either select a single prior from this set or adopt a prior averaging strategy. In the latter case, the final prior is formed by averaging over the set of plausible priors, thereby reflecting the uncertainty inherent in the prior specification process.

\subsection{Implementation}
The introduced methodological framework is implemented via the Python package \pkg{elicito}, which is released under the open-source Apache 2.0 license. \pkg{elicito} is available on PyPI (\url{https://pypi.org/project/elicito/}) and conda-forge (\url{https://anaconda.org/conda-forge/elicito}). The documentation is maintained on ReadTheDocs (\url{https://elicito.readthedocs.io}) and the codebase is hosted on GitHub (\url{https://github.com/florence-bockting/elicito}) to facilitate version control and collaborative development. The code adheres to the Black style guide, a widely adopted standard for Python formatting and linting. The package supports Python versions 3.9 through 3.12 and depends on \code{tensorflow >= 2.16}, \\\code{tensorflow-probability >= 0.24}, \code{tf-keras >= 2.16}, \code{numpy >= 1.24}, \code{joblib >= 1.4.2}, and \code{tqdm >= 4.38}. All dependencies are managed via the \code{uv} Python package manager.

The primary user interface of \pkg{elicito} is the \code{Elicit} class, through which the user can specify the entire elicitation procedure. The arguments of the \code{Elicit} class are designed to capture all necessary information required to implement an elicitation method. A brief overview of these arguments is provided below:
\begin{itemize}
    \item \code{model}: Defines the generative model used in the elicitation procedure.
    \item \code{parameters}: Specifies assumptions regarding the prior distributions over model parameters, including (hyper)parameter constraints, dimensionality, and parametric form.
    \item \code{targets}: Defines the elicited statistics in terms of target quantities and corresponding elicitation techniques. Specifies the discrepancy measure and weight used for the associated loss component.
    \item \code{expert}: Provides the expert information that serves as the basis for the learning criterion.
    \item \code{optimizer}: Specifies the optimization algorithm to be used, along with its hyperparameters (e.g., learning rate).
    \item \code{trainer}: Configures the overall training procedure, including settings such as the random seed, number of epochs, sample size, and batch size.
    \item \code{initializer}: Defines the initialization strategy for the hyperparameters used to instantiate the simulation-based optimization process; required only when using parametric prior distributions.
    \item \code{networks}: Specifies the architecture of the deep generative model; required only when using non-parametric prior distributions.
\end{itemize}
By configuring these core components, the \pkg{elicito} package supports a wide range of elicitation methods, including both structural and predictive approaches \citep{kadane1998experiences, falconer2022methods}. It accommodates univariate and multivariate as well as parametric and nonparametric prior distributions. In the following section, we detail the specification of each argument and demonstrate the full range of functionality offered by \pkg{elicito}.

\newpage
\section{Software}\label{sec:software}
The Python package \pkg{elicito} implements the simulation-based prior elicitation framework proposed by \citet{bockting2023simulation, bockting2024expert}. An overview of the underlying computational algorithm is provided in Section~\ref{subsec:computational-algorithm}. The subsequent sections introduce and explain the arguments of the \code{Elicit} class, which constitutes the primary user interface of \pkg{elicito}. In the following, we assume that the \pkg{elicito} package has been imported using the alias \code{el}.

\subsection{Computational algorithm}
\label{subsec:computational-algorithm}
Algorithm~\ref{alg:elicito-dag} outlines the computational steps underlying the \pkg{elicito} framework. In this pseudo-code we assume a single batch while we use a batch size greater one in the actual implementation, resulting in all computational objects carrying an additional dimension of size $B$.
\begin{algorithm}[h]
\caption{Computational algorithm underlying the \pkg{elicito} framework (single batch)}\label{alg:elicito-dag}
\begin{algorithmic}
    \Require $\lambda_0$
    \Comment{specify initialization for hyperparameter}
    \Require either \text{DNN} or $p_\lambda(\cdot)$
    \Comment{\parbox[t]{8cm}{specify non-parametric prior via deep generative model or parametric prior distribution families}}
    \Require $p\left(y\mid\theta\right)$
    \Comment{define generative model}
    \Require $c_i$ with $i=1,\ldots I$
    \Comment{define set of target quantities}
    \Require $f_j$ with $j=1,\ldots,J$
    \Comment{define elicitation techniques}
    \Require $D_m, w_m$ for $m=1,\ldots,M$
    \Comment{\parbox[t]{7cm}{define discrepancy measure and weight of individual loss component}}
    \Require epochs, $S$, $\eta$
    \Comment{\parbox[t]{8cm}{define training settings: learning rate for SGD optimizer, number of epochs and prior samples}}
    \vspace{1em}
    \State $\lambda \leftarrow \lambda_0$ 
    \Comment{initialize hyperparameter $\lambda$}
    \For{epoch in epochs}
        \State $p_\lambda \leftarrow \text{DNN}(\lambda)$
        \Comment{(optional) learn prior density from DNN}
        \For{$s = 1,\ldots,S$}
            \State draw $\theta^{(s)}$ from $p_\lambda(\cdot)$
            \Comment{sample model parameter from prior}
            \For{$n=1,\ldots,N$}
                \State draw $y_n^{(s)}$ from $p\left(y\mid\theta^{(s)}\right)\cdot p_\lambda\left(\theta^{(s)}\right)$
                \Comment{sample from generative model}
            \EndFor
            \For{$i=1,\ldots,I$}
                \State $\{z_i^{(s)}\} \leftarrow c_i(\lambda)$
                \Comment{compute target quantity}
            \EndFor
            \For{$m=1,\ldots,M$}
                \State $\{t_m^{(s)}\} \leftarrow f_j(z_i^{(s)})$
                \Comment{compute elicited statistics} 
                \State $L_m \leftarrow w_m D_m(t_m^{(s)}(\lambda), \hat t_m)$
                \Comment{compute individual loss component} 
            \EndFor
        \EndFor
    \State $\mathcal L = \sum_{m=1}^M L_m$
    \Comment{Compute total loss}
    \State $\lambda_\text{epoch+1} \leftarrow \lambda_\text{epoch}-\eta\nabla_{\lambda_\text{epoch}} \mathcal L$
    \Comment{Update $\lambda$ via backpropagation}
    \EndFor 
\end{algorithmic}
\end{algorithm}

\subsection{Setup Stage}
In the \emph{setup stage}, the generative model is defined, assumptions about the prior distributions are specified, and the set of target quantities to be elicited from the expert is identified, along with the corresponding elicitation techniques used to gather expert input. These configurations are specified through the arguments \code{model}, \code{parameters}, and \code{targets} of the \code{Elicit} class, which are detailed in the following subsections.

\subsubsection{Generative model}
The generative model is specified using the \code{model} argument of the \code{Elicit} class, via the \code{el.model()} function. This function follows the general structure
\begin{CodeChunk}
    \begin{CodeInput}
    model = el.model(obj=<class callable>, **kwargs).
    \end{CodeInput}
\end{CodeChunk}
The generative model must be implemented as a Python class that defines a \code{call} method. This method has to accept \code{prior_samples} as a mandatory input argument, with additional optional arguments as needed. It must return a dictionary containing the desired model-derived target quantities, where each entry corresponds to a name-value pair representing a specific target quantity. 

The class implementing the generative model is passed to the \code{el.model()} function via the \code{obj} argument. Any additional parameters required by the model can be specified using keyword arguments (\code{**kwargs}).

The following example illustrates an implementation for a simple normal regression model with design matrix $\mathbf{X}$, formally defined as
\begin{align*}
    \mu &= (\beta_0, \beta_1)^\top \mathbf{X}\\
    \textbf{y} &\sim \text{Normal}(\mu,\sigma).
\end{align*}
The implementation of the Python class is
\begin{CodeChunk}
    \begin{CodeInput}
    class GenerativeModel:
        def __call__(self, prior_samples, X):
            # extract samples per model parameters
            beta0, beta1, sigma = [prior_samples[:,:,i] for i in range(3)]
            
            # compute linear predictor term
            mu = beta0 + beta1*X
            
            # sample prior predictions from likelihood
            y = tfd.Normal(mu, sigma[:,:,None]).sample()

            return dict(y=y)
    \end{CodeInput}
\end{CodeChunk}
The corresponding model specification is \code{el.model(obj=GenerativeModel, X=X)}.

\paragraph{Special case: Discrete random variables}
When using mini-batch SGD as optimization procedure, it is essential that gradients can be computed at each step of the computational graph. However, this is generally infeasible for discrete random variables \citep[for a detailed discussion, see][]{bockting2023simulation}. One strategy to overcome this limitation involves approximating a discrete random variable with a continuous distribution, thereby enabling the use of gradient-based optimization methods. One such technique is the Gumbel-Softmax trick \citep{maddison2017concrete, jang2017categorical, joo2020generalized}, which is implemented in \pkg{elicito} via the \code{el.utils.gumbel_softmax_trick()} function. This function facilitates the definition of discrete likelihoods within the \code{GenerativeModel}, with current support limited to lower- and double-bounded random variables.

The function \code{el.utils.gumbel_softmax_trick()} requires two arguments: \code{likelihood} which expects a member of \code{TensorFlow Probability distributions (tfd)} representing the discrete likelihood and \code{upper_thres} which expects an integer defining the upper truncation threshold for lower-bounded distributions. For double-bounded distributions, this value corresponds to the upper bound.
An additional optional argument is the softmax temperature, \code{temp}, with default \code{temp=1.6}, a value that has shown good empirical performance.

The following example shows the implementation of a generative model with Binomial likelihood. To highlight the differences from the previous example using a continuous likelihood, each modified line is marked with \code{>}\code{>}:
\begin{CodeChunk}
    \begin{CodeInput}
       class GenerativeModel:
    >>      def __call__(self, prior_samples, X, total_count):
                # extract samples per model parameters
                beta0, beta1, sigma = [prior_samples[:,:,i] for i in range(3)]
            
                # compute linear predictor term
                mu = beta0 + beta1*X
                 
                # define discrete likelihood
    >>          likelihood = tfd.Binomial(total_count, tf.math.sigmoid(mu))
               
                # approximate samples from Binomial distribution
    >>          y = el.utils.gumbel_softmax_trick(
                        likelihood=likelihood, 
                        upper_thres=total_count
                    )

                return dict(y=y)
    \end{CodeInput}
\end{CodeChunk}
The corresponding model specification is then provided via \code{el.model(obj=GenerativeModel, X=X, total_count=total_count)}.

\subsubsection{Model parameters}\label{subsubsec: model-parameters}
Once the generative model has been defined, the assumptions regarding the prior distributions over the model parameters must be specified. This is accomplished via the \code{parameters} argument of the \code{Elicit} class.
This argument expects a list of model parameter specifications, with each parameter defined using the \code{el.parameter()} function. The required input format for \code{el.parameter()} depends on whether a parametric or non-parametric prior is assumed.

If a \emph{parametric prior} is assumed, both the distribution family and its associated hyperparameters must be provided. In this case, the \code{el.parameter()} function requires three arguments: \code{name}, \code{family}, and \code{hyperparams}. For instance, specifying a half-normal prior distribution for a scale parameter, $\sigma \sim \text{HalfNormal}(\sigma_0)$, is implemented as follows:
\begin{CodeChunk}
    \begin{CodeInput}
    el.parameter(
        name="sigma",
        family=tfd.HalfNormal, 
        hyperparams=dict(scale=el.hyper("sigma0", lower=0.))
    ).
    \end{CodeInput}
\end{CodeChunk}
The \code{name} argument assigns a custom identifier to the model parameter. The \code{family} argument specifies the prior distribution family, currently limited to members of the \code{tfd} module. Hyperparameters for the selected prior distribution family are provided via the \code{hyperparams} argument, which accepts a dictionary. Each key corresponds to a required input parameter of the specified \code{tfd} member, while the associated values are defined using the \code{el.hyper()} function. This function enables users to assign custom names to hyperparameters and impose constraints where applicable. 

The \code{el.hyper()} function supports multiple configurations, with the default configuration:
\begin{CodeChunk}
    \begin{CodeInput}
    el.hyper(name=<custom name>, lower=float("-inf"), upper=float("inf"), 
        vtype="real", dim=1, shared=False)
    \end{CodeInput}
\end{CodeChunk}
By default, a hyperparameter is treated as an unconstrained scalar. Constraints can be imposed via the \code{lower} and \code{upper} arguments to enforce lower bounds, upper bounds, or double-bounded intervals. The hyperparameter’s dimensionality is controlled using the \code{vtype} (variable type) and \code{dim} (dimension) arguments.
Additionally, hyperparameters can be \emph{shared} across multiple prior distributions by setting the argument \code{shared=True}. The following code snippet illustrates three examples: (1) an unconstrained hyperparameter, (2) a non-negative shared hyperparameter, and (3) a multidimensional hyperparameter:
\begin{CodeChunk}
    \begin{CodeInput}
    # unconstrained scalar hyperparameter
    el.hyper(name="mu0")
    
    # non-negative scalar hyperparameter shared across priors
    el.hyper(name="sigma0", lower=0., shared=True)
    
    # unconstrained 2D hyperparameter
    el.hyper(name="tau", vtype="array", dim=2)
    \end{CodeInput}
\end{CodeChunk}

When a \emph{non-parametric prior} is assumed for a model parameter, the \code{el.parameter()} function requires only the \code{name} argument. Additional constraints on the parameter’s domain can be specified using the \code{lower} and \code{upper} arguments. For example, \code{el.parameter(name="sigma", lower=0)} specifies a non-parametric prior for the scale parameter \code{sigma}, constrained to be non-negative.

Note that constraints on model parameters are only required when \emph{non-parametric} priors are specified. In the case of parametric priors, constraints on the \emph{hyper}parameters inherently impose the corresponding constraints on the model parameters.
All implemented constraints follow a standardized form closely following the parameter constraint implementation in \pkg{Stan} \citep{carpenter2017stan}. 
Specifically, for double-bounded variables, the inverse of the log-odds transformation is applied and for variables with either a lower or upper boundary, a softplus transformation is used. In this latter case, we deviate from the conventional exponential transform, as empirical results suggest that the softplus transformation provides greater numerical stability.

The complete specification of the \code{parameters} argument in the \code{Elicit} class for a generative model with two parameters, assuming a non-parametric joint prior, is illustrated in the following example:
\begin{CodeChunk}
    \begin{CodeInput}
    parameters = [
        el.parameter(name="beta"),
        el.parameter(name="sigma", lower=0)
        ]
    \end{CodeInput}
\end{CodeChunk}

\subsubsection{Target quantities and Elicitation techniques}
The specification of the target quantities and their associated elicitation techniques is provided via the \code{targets} argument of the \code{Elicit} class.
This argument accepts a list of elements, each defined using the \code{el.target()} function. The function follows the default configuration
\begin{CodeChunk}
    \begin{CodeInput}
    el.target(name=<custom name>, query=<el.queries>, loss=<function callable>, 
        target_method=None, weight=1.0).
    \end{CodeInput}
\end{CodeChunk}

\paragraph{Target quantities}
As introduced in Section~\ref{subsec:target-quantities-and-elicited-statistics}, a target quantity is defined as a function of the hyperparameters~$\lambda$, $z_i = c_i(\lambda)$. The transformation function $c_i$ can be implemented directly within the generative model, with the resulting target quantity $z_i$ returned as a key-value pair in the output dictionary. In this case, the corresponding target is specified using \code{el.target(name="z_i", target_method=None, ...)}. Here, \code{"z_i"} must match the key under which the generative model returns the computed quantity.

Alternatively, the transformation can be decoupled from the generative model by implementing a user-defined Python function. For example, consider the target quantity of interest is the coefficient of determination, defined as $R^2=\text{Var}(\mu)/\text{Var}(y)$ \citep{gelman2019r}. This transformation can be specified via a custom function, as illustrated below:
\begin{CodeChunk}
    \begin{CodeInput} 
    def r2(mu, y):
        return tf.divide(tf.math.reduce_variance(mu, -1),
                         tf.math.reduce_variance(y, -1)
                         )
    \end{CodeInput}
\end{CodeChunk}
In this example, the input arguments \code{mu} and \code{y} correspond to keys in the output dictionary returned by the generative model. The user-defined function is then passed to the \code{el.target()} function via the \code{target_method} argument: \code{el.target(name="r2_custom", target_method=r2, ...)}. Here, the \code{name} argument can be chosen freely. 

\paragraph{Elicitation techniques}
Once the target quantity is defined, the corresponding \emph{elicitation technique} can be specified via the \code{query} argument of the \code{el.target()} function. To support this process, several pre-implemented elicitation techniques are available through the \code{el.queries} class. Alternatively, users may define and supply their own elicitation technique via a custom function. The following examples illustrate the different approaches:
\begin{CodeChunk}
    \begin{CodeInput}
    # computes the quantiles of the provided target quantity  
    el.queries.quantiles(quantiles=(0.25, 0.50, 0.75))
    
    # returns the target quantity as is
    el.queries.identity()
    
    # computes the correlation between model parameters
    el.queries.correlation()

    # defines a custom elicitation technique
    def custom_func(target_name):
        return f(target_name)
        
    el.queries.custom(custom_func)
    \end{CodeInput}
\end{CodeChunk}
The \code{el.queries.quantiles()} function computes specified quantiles of the target quantity and requires as input a list or tuple of probabilities representing the desired quantile levels. The \code{el.queries.identity()} function returns the target quantity without applying any transformation, effectively serving as a pass-through operator. The \code{el.queries.correlation()} function computes the correlation between model parameters, which is particularly useful for learning joint priors in non-parametric settings. Lastly, the \code{el.queries.custom()} function allows users to define and incorporate custom elicitation techniques.

Note that although both target quantities and elicited statistics are defined within the \\ \code{el.target()} function, their computation is carried out sequentially (see Algorithm~\ref{alg:elicito-dag}). First, the target quantities are either extracted directly from the output dictionary of the generative model or computed via a user-defined transformation function. This results in a dictionary in which each entry corresponds to a distinct target quantity. In the second step, the elicitation technique, specified via the \code{query} argument, is applied to the corresponding target quantity. The output is a dictionary containing the simulated elicited statistics.

\paragraph{Discrepancy measure and weight}
Each simulated elicited statistic is compared to its corresponding expert-elicited counterpart using an appropriate discrepancy measure. Together with a weight $w_m$, this defines an individual loss component (see Section~\ref{subsec:expert-elicited-and-model-implied-statistics}). The discrepancy measure and its associated weight are specified via the \code{loss} and \code{weight} arguments in the \code{el.target()} function, respectively. The \pkg{elicito} package currently includes two built-in discrepancy measures: a squared-error loss (\code{el.losses.L2}) and a squared, biased Maximum Mean Discrepancy loss \citep[MMD,][]{gretton2012kernel}  (\code{el.losses.MMD2}), which supports both energy and Gaussian kernels. For greater flexibility, users may also provide custom discrepancy functions tailored to their specific application.

In summary, a complete specification of the \code{targets} argument within the \code{el.Elicit} class is illustrated in the following example:
\begin{CodeChunk}
    \begin{CodeInput}
    targets = [
        el.target(name="y", query=el.queries.quantiles((0.25, 0.50, 0.75)), 
            loss=MMD2(kernel="energy"), weight=1.),
        el.target(name="R2", query=el.queries.custom(sd), target_method=r2, 
            loss=L2, weight=10.)
        el.target(name="mu", query=el.queries.identity(), 
            loss=MMD2(kernel="gaussian", sigma=1.), weight=1.)
     ] 
    \end{CodeInput}
\end{CodeChunk}

\subsection{Elicitation stage}
Once the set of elicited statistics has been defined, the corresponding expert information must be collected and provided as input to the elicitation procedure. This is achieved through the \code{expert} argument of the \code{el.Elicit} class.

Expert-provided data is supplied to the elicitation method using the \code{el.expert.data()} function. This function takes a dictionary containing the expert information via its \code{dat} argument. The keys in this dictionary must exactly match a predefined structure that depends on the \code{targets} configuration. To facilitate this step, \pkg{elicito} offers the helper function \code{el.utils.get_expert_datformat()}, which generates a template dictionary illustrating the expected key-value structure for a given \code{targets} specification.

The following example illustrates a scenario in which an expert provides information about the expected outcome distribution, conditioned on each level of a three-level categorical predictor. A quantile-based elicitation approach is employed. 
\begin{CodeChunk}
    \begin{CodeInput}
    targets = [
        el.target(
            name=f"gr{i}",
            query=el.queries.quantiles((.05, .25, .50, .75, .95)),
            loss=el.losses.MMD2(kernel="energy"),
            weight=1.0) for i in [1,2,3]
            ]

    expert = el.expert.data(dat={
                "quantiles_gr1": [-12.55, -0.57, 3.29, 7.14, 19.15],
                "quantiles_gr2": [-11.18, 1.45, 5.06, 8.83, 20.42],
                "quantiles_gr3": [-9.28, 3.09, 6.83, 10.55, 23.29]
            })
    \end{CodeInput}
\end{CodeChunk}

\subsection{Fitting stage}
During the \emph{fitting} stage, the primary objective is to minimize the total loss function to learn the hyperparameters $\lambda$ that define the prior distributions aligned with the expert expectations. The configuration of the optimization procedure and the learning process is controlled through the \code{optimizer} and \code{trainer} arguments of the \code{Elicit} class. Additionally, when parametric priors are to be learned, the initial values required to instantiate the learning process must be specified using the \code{initializer} argument. In contrast, if non-parametric priors are used, the deep generative model needs to be configured through the \code{networks} argument. In the following sections, each of these arguments is described in detail.

\subsubsection{Optimization procedure}
The current optimization method supported in \pkg{elicito} is mini-batch SGD (see Section~\ref{subsec:prior-hyperparameter-optimizsation}). This requires the specification of an optimization algorithm, which can be provided via the \code{el.optimizer()} function and the corresponding argument of the \code{Elicit} class. The optimizer must be an instance of the \code{tf.keras.optimizers} module. By default, the Adam optimizer is used \citep{kingma2014auto}.

Optimizer-specific parameters, such as the learning rate, can be specified as additional keyword arguments to the \code{el.optimizer()} function. The learning rate may be provided either as a fixed scalar value or as a learning rate schedule, implemented as a member of the \code{tf.keras.optimizers.schedules} module.

The following example demonstrates the configuration of an Adam optimizer using a cosine decay learning rate schedule, along with an additional \code{clipnorm} argument, which constrains the gradient norm of each hyperparameter value $\lambda$.
\begin{CodeChunk}
    \begin{CodeInput}
    optimizer = el.optimizer(
        optimizer=tf.keras.optimizers.Adam,
        learning_rate=tf.keras.optimizers.schedules.CosineDecay(
            initial_learning_rate=0.1, decay_steps=600),
        clipnorm=1.0
        )
    \end{CodeInput}
\end{CodeChunk}

\subsubsection{Learning of prior distributions}
The overall learning process is configured via the \code{trainer} argument of the \code{Elicit} class, which is defined using the \code{el.trainer()} function. This function requires three arguments: \code{method}, \code{seed}, and \code{epochs}. The \code{method} argument specifies whether parametric (\code{"parametric\_prior"}) or non-parametric priors (\code{"deep\_prior"}) are assumed for the model parameters. The \code{seed} argument is applied internally to all sampling steps to ensure reproducibility. Finally, the \code{epochs} argument determines the number of iterations used to update the hyperparameters $\lambda$. 
In addition to the required arguments, \code{el.trainer} also accepts several optional arguments that further refine the training process. These include the batch size (\code{B=128}), the number of samples drawn from the prior distributions per epoch (\code{num_samples=200}), and the verbosity level (\code{progress=1}), which controls whether a progress bar is shown during training.

In the following example, we demonstrate the learning of parametric prior distributions over 400~epochs. In each epoch, 200 samples are drawn from the prior distributions to compute the model-implied elicited statistics. The batch size is set to 256, the random seed is 4, and progress output is disabled.
\begin{CodeChunk}
    \begin{CodeInput}
    trainer = el.trainer(
        method="parametric_prior",
        seed=4,
        epochs=400,
        B=256,
        num_samples=200,
        progress=0
    )
    \end{CodeInput}
\end{CodeChunk}

\subsubsection{Initialization and deep generative model}
To initiate simulation-based training, initial values for the hyperparameters $\lambda$ must be specified. For parametric priors, these initial values are provided via the \code{initializer} argument of the \code{Elicit} class. This is achieved using the \code{el.initializer()} function, which currently supports two initialization strategies: (1) explicitly specifying initial values for each hyperparameter $\lambda$, or (2) employing a method that explores the loss landscape to identify an initial configuration of $\lambda$ that yields a low loss value. 

For non-parametric priors, deep generative models are employed to represent the joint prior distribution. These models must be specified via the \code{network} argument of the \code{Elicit} class. Currently, \pkg{elicito} supports normalizing flows (NFs) for learning non-parametric joint priors. In this setting, the \code{initializer} argument should be set to \code{None}, as the deep generative model relies on default initialization schemes. Specifically, the kernel weight matrices in the dense layers are initialized using the Glorot uniform initializer, while the bias vectors are initialized to zero \citep{glorot2010understanding}.

\subsection{Evaluation stage}
\subsubsection{Instantiation and methods of the Elicit class}
Once all arguments of the \code{Elicit} class have been specified, an instance of the class can be created. We refer to this instance as \code{eliobj} in the following.
\begin{CodeChunk}
    \begin{CodeInput}
    eliobj = el.Elicit(
        model=model,
        parameters=parameters,
        targets=targets,
        expert=expert,
        optimizer=optimizer,
        trainer=trainer,
        initializer=initializer,
        networks=None
    )
    \end{CodeInput}
\end{CodeChunk}
Each argument passed to the \code{Elicit} class is stored as a corresponding attribute within the \code{eliobj} instance, enabling retrieval and inspection of the configuration at a later stage. In addition, two attributes, \code{history} and \code{results}, are initialized as empty lists upon instantiation. These serve as containers for recording results from the fitting process during optimization. The fitting procedure is initiated using the \code{eliobj.fit()} method.

The \code{fit()} method supports parallel execution via its \code{parallel} argument, which enables running multiple fitting processes with different random seeds simultaneously. This parallelization is configured through the \code{el.parallel()} function.

An \code{Elicit} instance can be saved to disk using the \code{eliobj.save()} method and subsequently restored using the \code{el.utils.load()} helper function. Furthermore, the \code{eliobj.update()} method facilitates targeted modifications to one or more arguments of the original \code{Elicit} configuration. This allows users to selectively update components without the need to re-specify the entire configuration.

\subsubsection{Postprocessing and evaluation of optimisation results}
The outcomes of the fitting procedure are stored in the \code{history} and \code{results} attributes of the \code{eliobj} instance. When multiple replications are performed in parallel, the results corresponding to each replication are stored in separate sublists, resulting in both \code{history} and \code{results} being represented as nested lists. Results from a specific replication can be accessed using standard Python indexing; for instance, \code{eliobj["results"][2]} retrieves the results from the third replication.

The \code{history} attribute records detailed information at each iteration across all epochs, including quantities such as loss values, hyperparameter values, and iteration durations. These data are particularly useful for monitoring the training progress and conducting convergence diagnostics.

In contrast, the \code{results} attribute stores results only once, upon completion of the fitting process. These include, for example, simulated prior samples, samples drawn from the generative model, and simulated elicited statistics. Such outputs are valuable for evaluating the learned prior distributions and for analyzing the model-implied quantities that result from these priors.

To facilitate result inspection, \pkg{elicito} provides a set of built-in graphical evaluation tools implemented in the \code{el.plots} module.
Among others, \code{el.plots.loss()} and \\ \code{el.plots.hyperparameter()} allow for visualization of the learning progress over time, \\ \code{el.plots.elicits()} enables comparison between model-implied and expert-elicited statistics, and \code{el.plots.prior_joint()}, \code{el.plots.prior_marginals()}, and \\ \code{el.plots.prior_averaging()} allow for examining the learned prior distributions. 

\section{Case Study}
\label{sec:case-study}
In this section, we illustrate the use of the \pkg{elicito} package through two toy examples. We begin with a scenario in which independent parametric prior distributions are assumed, and subsequently discuss the modifications required to accommodate a non-parametric joint prior distribution. The complete implementation of each toy example is provided in the supplementary material hosted on GitHub at \url{https://github.com/florence-bockting/elicito-software-paper}. 

\subsection{Example 1: Independent parametric priors}
For the following case study, we consider a linear regression model with a normally distributed likelihood and a three-level, dummy-coded categorical predictor. The corresponding statistical model ($\mathcal{M}$) comprises four parameters, $\boldsymbol\theta$: the intercept ($\beta_0$), two contrast coefficients associated with the categorical predictor ($\beta_1$ and $\beta_2$), and the residual standard deviation~($\sigma$). Formally, the model $\mathcal{M}$ is defined as
\begin{align*}\tag{$\mathcal{M}$} 
\begin{split} 
    \mu_i &= \beta_0 + \beta_1X_1+\beta_2X_2\\ 
    y_i &\sim \text{Normal}(\mu_i, \sigma) 
\end{split} 
\end{align*}
where the observations $y_i$ are drawn from a normal distribution centered at the predicted mean $\mu_i$. 
In this first example, we assume independent parametric prior distributions for the model parameters $\boldsymbol{\theta}$. Specifically, the regression coefficients are assigned normal priors, while the residual standard deviation $\sigma$ is assigned a half-normal prior:
\begin{align*}\tag{$\mathcal P_\text{parametric}$}
\begin{split}
    \beta_j &\sim \text{Normal}(\mu_j, \sigma_j) \quad \text{for }j = 0,1,2\\ 
    \sigma &\sim \text{HalfNormal}(\sigma_3).
\end{split}
\end{align*}
Under this specification, the objective is to learn a set of seven hyperparameters, denoted by $\lambda=(\mu_0, \sigma_0, \mu_1, \sigma_1, \mu_2, \sigma_2, \sigma_3)$.

Next, we define the set of quantities to be elicited from the domain expert. We assume that the expert is able to express expectations regarding the dependent variable’s outcome within each of the three groups, denoted as $y_i \mid \text{gr}_k$ for $k = 1, 2, 3$. In addition, we include the expected coefficient of determination ($R^2$), representing the proportion of variance explained by the predictor. Together, these four quantities form the set of \emph{target quantities}, denoted as ${z_i} = (z_1, z_2, z_3, z_4)$. 
To elicit this information from a domain expert, we employ a quantile-based elicitation approach for all four target quantities. Specifically, the expert is asked to provide the 25th, 50th, and 75th percentiles, denoted as $Q_{25}$, $Q_{50}$, and $Q_{75}$, for each target quantity. This procedure yields a set of four \emph{elicited statistics}, denoted as $\{t_m\} = (t_1, t_2, t_3, t_4)$ where
\begin{align*}
    t_{1} &= (Q_{25}(z_1), Q_{50}(z_1), Q_{75}(z_1)) \quad \text{with }z_1=y_i \mid \text{gr}_1, \\ 
    &\ldots \\
    t_{4} &= (Q_{25}(z_4), Q_{50}(z_4), Q_{75}(z_4)) \quad \text{with }z_4=R^2.
\end{align*}
To estimate the seven hyperparameters $\lambda$, the optimization procedure searches for values that minimize the discrepancy between the expert-elicited and model-simulated sets of elicited statistics, ${t_m}$. Based on this setup, we now describe how to specify the corresponding \code{Elicit} class using the \pkg{elicito} package, assuming it has been imported in Python via the alias \code{el}.

\subsubsection{Generative model}
Information about the generative model is specified via the \code{model} argument of the \code{Elicit} class. In the current example, the data-generating process is implemented through the \code{GenerativeModel} class, which returns information about the specified target quantities $z_i$.
\begin{CodeChunk}
    \begin{CodeInput}
    class GenerativeModel:
        def __call__(self, prior_samples, design_matrix, n_gr):
            # extract prior samples per parameter type
            betas=prior_samples[:,:,:-1]
            sigma=prior_samples[:,:,-1][:,:,None]
            
            # linear predictor
            mu = tf.matmul(betas, design_matrix, transpose_b=True)
            
            # data-generating model
            y = tfd.Normal(loc=mu, scale=sigma).sample()
            
            # selected observations per group
            (y_gr0, y_gr1, y_gr2) = (y[:, :, i::] for i,j in zip(
                                    [0,n_gr,2*n_gr],[n_gr,2*n_gr,-1]))
                                    
            return dict(y_gr0=y_gr0, y_gr1=y_gr1, y_gr2=y_gr2,
                        mu=mu, y=y)
    \end{CodeInput}
\end{CodeChunk}
The corresponding \code{model} argument of the \code{Elicit} class is then specified as follows:
\begin{CodeChunk}
    \begin{CodeInput}
    model=el.model(
        obj=GenerativeModel,
        n_gr=30,
        design_matrix=design_categorical(n_gr=30)
      )
    \end{CodeInput}
\end{CodeChunk}
The design matrix is constructed using the user-defined function \code{design_categorical()}, whose implementation is provided in Appendix~\ref{app: impl-param-model}.

\subsubsection{Prior specifications}
Next, the prior assumptions $\mathcal{P}_\text{parametric}$ are incorporated by specifying the \code{parameters} argument of the \code{Elicit} class. In this example, we assume independent parametric priors for each of the four model parameters $\theta$.\footnote{For readers unfamiliar with Python syntax, the expression \code{[x for x in range(3)]} represents a \emph{list comprehension}. This approach offers a more compact alternative to traditional for-loops. Furthermore, the addition of two lists using the \code{+} operator (i.e., \code{list() + list()}) results in their concatenation.}
\begin{CodeChunk}
    \begin{CodeInput}   
    parameters=[
            el.parameter(
              name=f"beta{i}",
              family=tfd.Normal,
              hyperparams=dict(loc=el.hyper(f"mu{i}"),
                               scale=el.hyper(f"sigma{i}", lower=0))
            ) for i in range(3)
          ]+[
            el.parameter(
              name="sigma",
              family=tfd.HalfNormal,
              hyperparams=dict(scale=el.hyper("sigma3", lower=0))
            ),
          ]
    \end{CodeInput}
\end{CodeChunk}
In this implementation, constraints are applied to each scale hyperparameter $\sigma_i$ by setting \code{lower=0}, thereby ensuring non-negativity.

\subsubsection{Target quantities and elicitation techniques}
Information about each of the four target quantities $z_i$ with respective elicitation technique is provided through the \code{targets} argument of the \code{Elicit} class. 
\begin{CodeChunk}
    \begin{CodeInput} 
    targets=[
        el.target(
          name=f"y_gr{i}",
          query=el.queries.quantiles(quantiles=(.25, .50, .75)),
          loss=el.losses.MMD2(kernel="energy"),
        ) for i in range(3)
        ]+[
        el.target(
          name="r2",
          query=el.queries.quantiles(quantiles=(.25, .50, .75)),
          loss=el.losses.MMD2(kernel="energy"),
          target_method=r2,
          weights=10.
        )
      ]
    \end{CodeInput}
\end{CodeChunk}
The target quantities $y_i \mid \text{gr}_k$ for $k = 1, 2, 3$ are extracted directly from the output dictionary of the \code{GenerativeModel}, while the target quantity $R^2$ is computed using a custom function that takes the returned values \code{y} and \code{mu} from the \code{GenerativeModel} as input parameters.
\begin{CodeChunk}
    \begin{CodeInput} 
    def r2(y, mu):
      return tf.divide(
        tf.math.reduce_variance(mu, axis=-1),
        tf.math.reduce_variance(y, axis=-1)
      )
    \end{CodeInput}
\end{CodeChunk}
The discrepancy between each model-implied statistic $t_m$ and its corresponding expert-elicited counterpart $\hat{t}_m$ is quantified using the $\text{MMD}^2$ loss with an energy kernel. The loss weights are set to unity ($w_1 = w_2 = w_3 = 1$) for the first three statistics, while a higher weight is assigned to $R^2$ ($w_4 = 10$).

\subsubsection{Expert data}
Having specified the generative model and the set of target quantities as well as elicited statistics, the corresponding quantities can be elicited from a domain expert. For the current example, we consider the following expert data as input:
\begin{CodeChunk}
    \begin{CodeInput}
    expert = el.expert.data(
        dat={'quantiles_y_gr0': [0.64, 1.22, 1.89], 
             'quantiles_y_gr1': [0.72, 1.39, 2.07], 
             'quantiles_y_gr2': [0.76, 1.48, 2.22], 
             'quantiles_r2':    [0.07, 0.23, 0.61]}
        )
    \end{CodeInput}
\end{CodeChunk}

\subsubsection{Optimization procedure}
The remaining arguments of the \code{Elicit} class (i.e., \code{optimizer}, \code{trainer}, and \code{initializer}) are used to configure the optimization procedure itself. For the current example, we employ mini-batch SGD with the Adam optimizer and a constant learning rate of 0.05, specified via \code{el.optimizer()} in the \code{optimizer} argument of \code{Elicit}:
\begin{CodeChunk}
    \begin{CodeInput}
    optimizer=el.optimizer(
        optimizer=tf.keras.optimizers.Adam,
        learning_rate=0.05
    )
    \end{CodeInput}
\end{CodeChunk}
Additional configurations are passed via the \code{trainer} argument of the \code{Elicit} class. These include the number of training epochs, the seed for reproducibility, and the specification of which approach is used to learn the prior distributions. In the present example, we employ the \code{"parametric_prior"} approach, which does not utilize a deep generative model to learn the priors:
\begin{CodeChunk}
    \begin{CodeInput}
    trainer=el.trainer(
        method="parametric_prior",
        seed=1234,
        epochs=600
    )
    \end{CodeInput}
\end{CodeChunk}
Finally, the initial hyperparameters $\boldsymbol\lambda^0$ must be specified to initiate the simulation-based training procedure. In the following example, 32 hyperparameter vectors are sampled from a $\text{Mv-Uniform}(0,2)$ distribution using a quasi-random sampling approach (i.e., Sobol sampling). For each sampled hyperparameter vector, the corresponding loss value is computed by running the elicitation method in forward mode. The hyperparameter vector $\boldsymbol{\lambda}^*$ that yields the minimum loss is selected as the set of initial values $\boldsymbol{\lambda}^0$:
\begin{CodeChunk}
    \begin{CodeInput}
    initializer=el.initializer(
        method="sobol",
        iterations=32,
        distribution=el.initialization.uniform(radius=2., mean=0.)
    )
    \end{CodeInput}
\end{CodeChunk}
With these settings, the \code{Elicit} class is fully specified and can be used to instantiate an \code{eliobj} object:
\begin{CodeChunk}
    \begin{CodeInput}
    eliobj = el.Elicit(
        model=model,
        parameters=parameters,
        targets=targets,
        expert=expert,
        optimizer=optimizer,
        trainer=trainer,
        initializer=initializer
    )
    \end{CodeInput}
\end{CodeChunk}

\subsubsection{Fitting and inspection of results}
In the current example, we run five instances of the optimization procedure in parallel to evaluate the sensitivity of the results to different initializations. 
\begin{CodeChunk}
    \begin{CodeInput}
    eliobj.fit(parallel=el.utils.parallel(runs=5))
    \end{CodeInput}
\end{CodeChunk}

\paragraph{Inspecting results}
The results of the optimization process are stored in \code{eliobj.results} and \code{eliobj.history}, both of which are nested lists, where each sublist corresponds to a single replication. The following example demonstrates how to inspect the results from the first optimization run.
\begin{CodeChunk}
    \begin{CodeInput}
    > eliobj.results[0].keys()

    dict_keys(['target_quantities', 'elicited_statistics', 'prior_samples', 
        'model_samples', 'loss_tensor_expert', 'loss_tensor_model', 
        'expert_elicited_statistics', 'expert_prior_samples', 
        'init_loss_list', 'init_prior', 'init_matrix', 'seed'])

    > eliobj.history[0].keys()

    dict_keys(['loss', 'loss_component', 'time', 'hyperparameter', 
        'hyperparameter_gradient'])
    \end{CodeInput}
\end{CodeChunk}

\paragraph{Convergence analyses}
The training progress can be examined using the graphical tools provided by the \code{el.plots} module. First, we visualize the trajectory of the total loss and the individual loss components, each corresponding to one of the elicited statistics, across epochs. Since multiple optimization runs were conducted in parallel, the results from all runs are displayed simultaneously in Figure~\ref{fig:plot-loss}.
\begin{CodeChunk}
    \begin{CodeInput}
        el.plots.loss(eliobj)
    \end{CodeInput}
\begin{figure}[ht]
    \centering
    \includegraphics[width=0.8\linewidth]{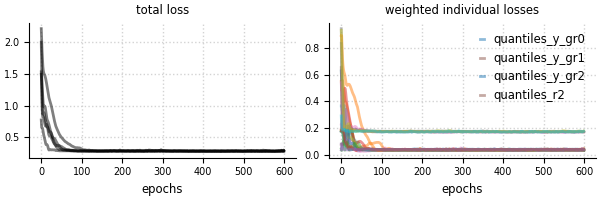}
    \caption{Convergence of total loss (left panel) and loss per component (right panel) across epochs for all parallel runs. Plot is created using \code{el.plots.loss(eliobj)}. The plots demonstrate successful convergence, with loss trajectories stabilizing at values near zero.}
    \label{fig:plot-loss}
\end{figure}
\end{CodeChunk}
In addition to evaluating the convergence of the loss components, it is informative to examine the convergence behavior of the individual hyperparameters $\lambda$ which is shown in Figure~\ref{fig:plot-hyperparameter}.
\begin{CodeChunk}
    \begin{CodeInput}
        el.plots.hyperparameter(eliobj)
    \end{CodeInput}
\begin{figure}
    \centering
    \includegraphics[width=0.8\linewidth]{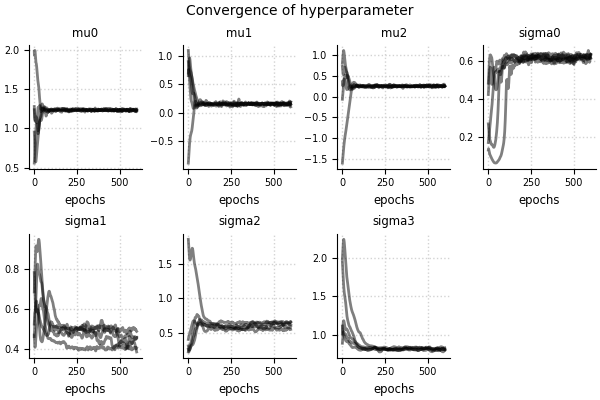}
    \caption{Convergence of hyperparameters $\lambda$ across epochs for all parallel runs. Plot is created using \code{el.plots.hyperparameter(eliobj)}. The learning trajectory of almost all hyperparameters stabilizes over time, with parallel runs converging towards the same value. Some difficulties of the optimization algorithm in learning $\sigma_1$ can be observed.}
    \label{fig:plot-hyperparameter}
\end{figure}
\end{CodeChunk}
Finally, we can also inspect the agreement between the expert-elicited statistics, $\hat{t}_m$, and the model-generated statistics, $t_m$ as visualized in Figure~\ref{fig:plot-elicits}.
\begin{CodeChunk}
    \begin{CodeInput}
        el.plots.elicits(eliobj)
    \end{CodeInput}
\begin{figure}[ht]
    \centering
    \includegraphics[width=0.8\linewidth]{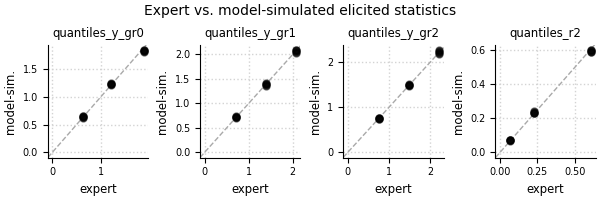}
    \caption{Comparison of expert-elicited and model-simulated statistics for all parallel runs. Plot is created using \code{el.plots.elicits(eliobj)}. For the current example, a perfect match is observed between the model-simulated and expert-elicited statistics for all four elicited statistics, as indicated by the scatter points lying along the diagonal line.}
    \label{fig:plot-elicits}
\end{figure}
\end{CodeChunk}
The gray dashed diagonal line in Figure~\ref{fig:plot-elicits} represents perfect agreement between model-implied and expert-elicited statistics. Deviations of the scatter points from this diagonal indicate discrepancies between the expert-elicited values and the statistics derived from the learned priors.

\paragraph{Learned prior distributions}
Upon confirming the successful learning of the optimization algorithm across all parallel runs, we proceed with examining the learned prior distributions corresponding to the elicited statistics. With \code{el.plots.prior_marginals(eliobj)} we can plot the learned marginal prior distributions for each model parameter as depicted in Figure~\ref{fig:plot-prior-marginal}.
\begin{CodeChunk}
    \begin{CodeInput}
        el.plots.prior_marginals(eliobj)
    \end{CodeInput}
\begin{figure}[ht]
    \centering
    \includegraphics[width=0.8\linewidth]{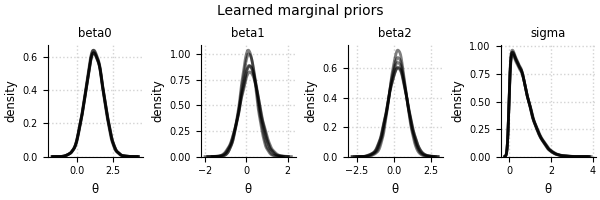}
    \caption{Learned marginal prior distribution for all parallel runs. The plot is generated using \code{el.plots.prior\_marginal(eliobj)}. }
    \label{fig:plot-prior-marginal}
\end{figure}
\end{CodeChunk}
Finally, a prior average across replications can be computed using \\\code{el.plots.prior_averaging(eliobj)} with results depicted in Figure~\ref{fig:plot-prior-average}. However, in the current example, this analysis offers limited additional insight, as the replications exhibit minimal variation.
\begin{CodeChunk}
    \begin{CodeInput}
        el.plots.prior_averaging(eliobj)
    \end{CodeInput}
\begin{figure}[ht]
    \centering
    \includegraphics[width=0.8\linewidth]{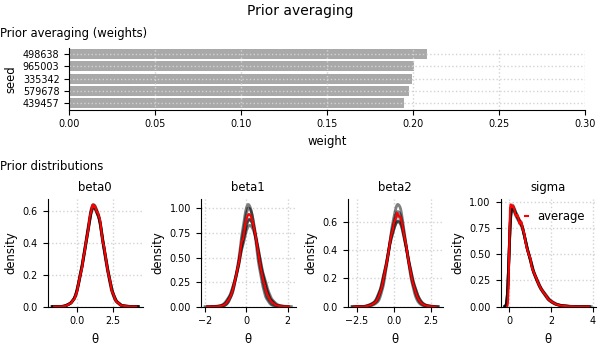}
    \caption{Prior averaging across multiple parallel runs with average prior highlighted in red. Upper panel shows weight associated with each replication used in model averaging. Lower plots shows the individual marginal prior distributions as well as the averaged prior distribution. Plot is created using \code{el.plots.prior\_averaging(eliobj)}.}
    \label{fig:plot-prior-average}
\end{figure}
\end{CodeChunk}
The upper panel in Figure~\ref{fig:plot-prior-average} displays the weights assigned to each replication during the averaging procedure, while the lower panel shows the corresponding marginal prior distributions. The averaged prior distribution is highlighted in red.

\subsection{Example 2: Joint non-parametric prior}
In the remainder of this section, we retain the previously introduced regression model $\mathcal{M}$ but change the assumptions regarding the prior distributions of the model parameters. Specifically, we assume a joint, non-parametric prior distribution of the form
\begin{align*}\tag{$\mathcal P_\text{non-parametric}$}
    (\beta_0, \beta_1, \beta_2, \sigma) \sim p_\lambda(\cdot).
\end{align*}
The hyperparameters $\lambda$ represent now the weights of the deep generative model used to learn the complex joint prior density function.
In the following, we revisit the specification of the \code{Elicit} class, focusing exclusively on aspects that differ from the previously discussed parametric case. For clarity, certain code-snippets from the previous example are repeated. In such instances, the symbol \code{>}\code{>} is used to indicate where exactly modifications relative to the previous example were made.

\paragraph{Prior specifications} 
The modified assumptions about the model parameters, $\mathcal{P}_\text{non-parametric}$, are incorporated through the \code{parameters} argument of the \code{Elicit} class. In this case, only the specification of the parameter names and their constraints is required.
\begin{CodeChunk}   
    \begin{CodeInput}
    parameters_deep=[
            el.parameter(name=f"beta{i}") for i in range(3)
        ]+[
            el.parameter(name="sigma", lower=0),
        ]
    \end{CodeInput}
\end{CodeChunk}

\paragraph{Elicited statistics and expert data}
The same expert information as described in the previous example is used in this case. Additionally, the independence assumption among the model parameters is explicitly incorporated as a loss component by specifying an additional \code{el.target()} function. The \code{targets} and \code{expert} arguments of the \code{Elicit} class are specified as follows:
\begin{CodeChunk}
    \begin{CodeInput}
    targets_deep = targets + [
    >>    el.target(
            name="cor",
            query=el.queries.correlation(),
            loss=el.losses.L2,
            weight=0.1
        )
    ]

    expert_deep = el.expert.data(
        dat={'quantiles_y_gr0': [0.64, 1.22, 1.89],
             'quantiles_y_gr1': [0.72, 1.39, 2.07],
             'quantiles_y_gr2': [0.76, 1.48, 2.22],
             'quantiles_r2': [0.07, 0.23, 0.61],
    >>       'cor_cor': [0.]*6
         }
    )
    \end{CodeInput}
\end{CodeChunk}
To quantify the deviation of the simulation-based correlations from zero (model parameters are assumed to be independent), the $L_2$ loss is used with an associated weight of $0.1$.

\paragraph{Optimization procedure}
The other adjustments refer to the optimization procedure itself (i.e., \code{optimizer}, \code{trainer}, and \code{network}).
We use again mini-batch SGD as optimization algorithm, but for the deep generative model approach typically smaller learning rates are required
\begin{CodeChunk}
    \begin{CodeInput}
      optimizer_deep=el.optimizer(
          optimizer=tf.keras.optimizers.Adam,
    >>    learning_rate=0.001
      )
    \end{CodeInput}
\end{CodeChunk}
For training, it is now necessary to specify that a deep generative model should be used to learn the prior, which is achieved by setting the approach to \code{"deep_prior"}. Additionally, the number of training epochs is increased, as deep generative models typically require more iterations to converge.
\begin{CodeChunk}
    \begin{CodeInput}
      trainer_deep=el.trainer(
    >>    method="deep_prior",
          seed=2025,
    >>    epochs=800
      )
    \end{CodeInput}
\end{CodeChunk}
Finally, we have to setup the deep-generative model via the \code{network} argument of the \code{Elicit} class. For the current example we use NFs implement via \code{el.networks.NF()}. We employ a relatively simple NF architecture, consisting of a standard multivariate Gaussian as base distribution and three affine coupling blocks. Each coupling block comprises two dense layers with 128 units and ReLU activation functions:
\begin{CodeChunk}
    \begin{CodeInput}
    network=el.networks.NF(
        inference_network=el.networks.InvertibleNetwork,
        network_specs=dict(
            num_params=4,
            num_coupling_layers=3,
            coupling_design="affine",
            coupling_settings={
                "dense_args": {
                    "units": 128,
                    "activation": "relu"
                },
                "num_dense": 2,
            },
        ),
        base_distribution=el.networks.base_normal
    )
    \end{CodeInput}
\end{CodeChunk}
With these modifications in place, we are ready to execute the elicitation procedure under the non-parametric prior assumption. Rather than reinitializing the \code{Elicit} class and creating a new \code{eliobj} instance, we update the existing \code{eliobj} instance created in the previous example and proceed to fit the updated object accordingly. 
\begin{CodeChunk}
    \begin{CodeInput}
    import copy
    
    # copy existing eliobj
    eliobj_deep = copy.deepcopy(eliobj)

    # update eliobj
    eliobj_deep.update(
        parameters=parameters_deep,
        targets=targets_deep,
        optimizer=optimizer_deep,
        trainer=trainer_deep,
        initializer=None,
        network=network
        )

    # fit updated eliobj
    eliobj_deep.fit(parallel=el.utils.parallel(runs=5))
    \end{CodeInput}
\end{CodeChunk}

\subsubsection{Fitting and inspection of results}
The fitting procedure and evaluation tools remain consistent with those described previously, with only minor modifications. In the following, we focus exclusively on the differences; the remaining plots are provided in Appendix~\ref{app: results-toy-example-2}.

The function \code{el.plots.hyperparameter(eliobj)} visualizes the convergence of the mean and standard deviation for each marginal prior distribution instead of the hyperparameter values $\lambda$, as depicted in Figure~\ref{fig:hyperparameter-deep-prior}.
\begin{CodeChunk}
    \begin{CodeInput}
        el.plots.hyperparameter(eliobj_deep)
    \end{CodeInput}
    \begin{figure}[h]
        \centering
        \includegraphics[width=0.8\linewidth]{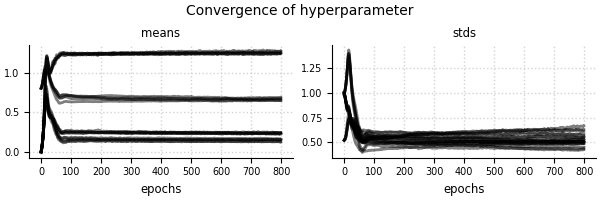}
        \caption{Convergence of mean and standard deviation of marginal prior distributions for all parallel runs. Plot is created using \code{el.plots.hyperparameter(eliobj\_deep)}.}
        \label{fig:hyperparameter-deep-prior}
    \end{figure}
\end{CodeChunk}
While the model reliably learns the locations of the prior distributions between the parallel runs, substantially greater variability is observed in the learned scales. This suggests that the information provided as input is insufficient to accurately inform the scale of the prior distributions, resulting in a lack of identifiability. 

The comparison between the simulated and expert-elicited statistics is shown in Figure~\ref{fig:elicits-deep-prior} and includes now also the correlation information. 
\begin{CodeChunk}
    \begin{CodeInput}
        el.plots.elicits(eliobj_deep)
    \end{CodeInput}
    \begin{figure}[h]
        \centering
        \includegraphics[width=0.8\linewidth]{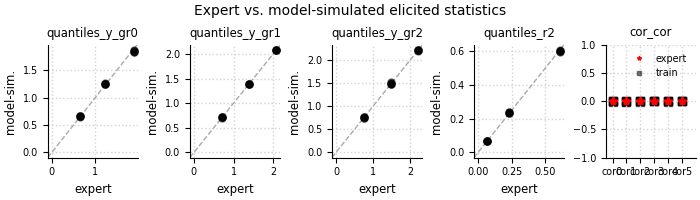}
        \caption{Comparison between simulated and expert-elicited statistics. Plot is created using \code{el.plots.elicits(eliobj\_deep)}. The first four plots (from the left) compare the model-implied and expert-elicited quantiles, while the right-most plot illustrates the independence assumptions among the model parameters, quantified via pairwise correlations. For all elicited statistics the expected and learned values exhibit a perfect match.}
        \label{fig:elicits-deep-prior}
    \end{figure}
\end{CodeChunk}
Finally, we can inspect the learned joint prior distribution using \\\code{el.plots.prior_joint(eliobj_deep, idx=list(range(5)))} depicted in Figure~\ref{fig:prior-joint-deep}.
\begin{CodeChunk}
    \begin{CodeInput}
        el.plots.prior_joint(eliobj_deep, idx=list(range(5)))
    \end{CodeInput}
    \begin{figure}[h]
        \centering
        \includegraphics[width=0.7\linewidth]{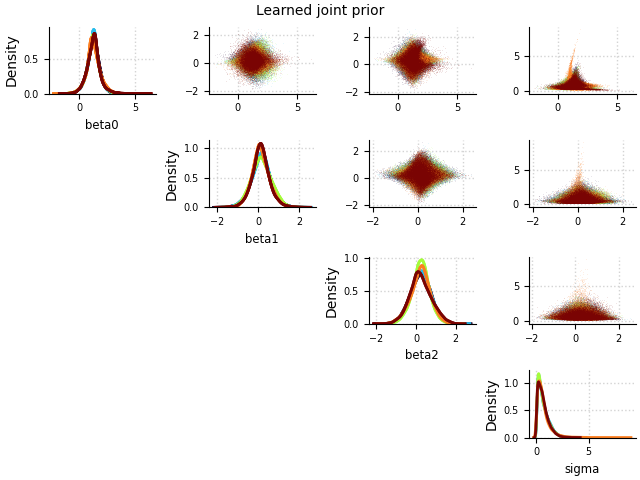}
        \caption{Learned prior distributions including information about correlation structure. The plot is created using \code{el.plots.prior\_joint(eliobj\_deep, idx=list(range(5)))}.}. 
        \label{fig:prior-joint-deep}
    \end{figure}
\end{CodeChunk}

\section{Related software}\label{sec:relevant-software}
Over the past decades, numerous prior elicitation methods have been proposed, many accompanied by specialized software implementations. Early examples of such tools, though no longer actively maintained, include \pkg{ElicitN} for expert knowledge elicitation on species richness and related count data \citep{fisher2012elicitn} or \pkg{Elicitator} for regression analysis in ecological applications \citep{james2010elicitator}. 

More recent examples, include among others the R package \pkg{DTEAssurance} \citep{salsbury2024assurance}, which implements the assurance method for delayed treatment effects by eliciting prior distributions for both the delay duration and post-delay hazard ratio. Similarly, the \pkg{assurance} package \citep{alhussain2020assurance} provides an implementation for normally distributed data, incorporating elicited prior distributions for treatment effects along with population variances in both treatment and control arms. More general approaches are offered for example via the \pkg{eglm} package \citep{hosack2024eglm}, which enables the elicitation of prior distributions for Bayesian generalized linear models and the \pkg{makemyprior} package \citep{hem2024makemyprior} for constructing joint priors for variance parameters in latent Gaussian models.

Despite these developments, general-purpose packages for expert prior elicitation remain relatively scarce \citep{mikkola2023prior}. Below, we present a selection of actively maintained tools that we consider most relevant to our current application.

The R package \pkg{pbbo} \citep{manderson2023translating} implements a simulation-based prior elicitation method using multi-objective Bayesian optimization. It enables elicitation based on expert-specified prior predictive distributions and model-derived quantities such as $R^2$. The Python package \pkg{PreliZ} \citep{icazatti2023preliz} provides a toolbox for elicitation and exploration of prior distributions. The package supports both structural and predictive elicitation through four core components: prior distribution specification and manipulation, visualization of priors and their predictive distributions, interactive user interfaces, and algorithms for prior inference from user-provided information. The R package \pkg{SHELF} \citep{oakley2025shelf} provides a comprehensive toolbox for prior elicitation within the Sheffield Elicitation Framework. It facilitates structured elicitation protocols involving single or multiple experts by offering multiple univariate elicitation methods (including roulette, bisection, and range techniques), support for multivariate parameter elicitation (such as vectors of proportions), and distribution fitting via least-squares minimization. It incorporates interactive feedback tools during elicitation sessions and is designed primarily as a facilitator's toolkit to support the Sheffield elicitation protocol. Finally, the R package \pkg{qdp} \citep{perepolkin2023quantile} introduces quantile and quantile-parameterized distributions as a flexible and interpretable approach to prior specification. These distributions allow expert knowledge to be expressed directly in the observable space, enhancing transparency and interpretability. The associated R package provides tools for defining and inverting quantile functions. 

Each of these packages adopts a distinct perspective and can be viewed as complementary to \pkg{elicito}. The aim of our package is to propose a unified workflow for prior elicitation, structured around a fixed set of interconnected modules, corresponding to the arguments of the \code{Elicit} class. The functionality of each module is designed to be extensible, allowing for integration with other packages where appropriate. For example, future versions of \pkg{elicito} could incorporate Bayesian optimization techniques, such as those implemented in \pkg{pbbo}. Similarly, the loss regularization methods proposed by \citet{manderson2023translating} represent a promising avenue for enhancing the loss module within \pkg{elicito}. The \pkg{PreliZ} package also complements \pkg{elicito} by offering exploratory tools that can enrich both the setup and evaluation stages of the elicitation process. Additionally, future extensions may include native support for quantile and quantile-parameterized priors, as implemented in \pkg{qdp}.

\section{Discussion and Conclusion}
\label{sec:discussion-and-conclusion}
The Python package \pkg{elicito} implements a modular workflow that encompasses the essential steps of the expert prior elicitation process. It provides a structured approach to the challenging task of formulating prior distributions that accurately reflect expert knowledge. To address this task, we use a simulation-based approach: (1) knowledge about desired quantities (target quantities) is extracted from a domain expert, (2) a forward model comprising of the generative model and the computation of target quantities, simulates the expert knowledge based on a set of parameters, (3) an optimization algorithm adjusts parameters to minimize the discrepancy between the simulated and observed expert knowledge; and (4) the resulting set of parameters defines the model parameter priors corresponding to the expert knowledge. Currently, \pkg{elicito} addresses this optimization task using SGD. Future developments should explore alternative optimization strategies, such as Bayesian optimization \citep[see for example,][]{manderson2023translating}, to enable comparative performance assessment of the resulting solutions.

Due to its modular structure and the resulting flexibility, \pkg{elicito} facilitates the implementation of a wide variety of prior elicitation methods.
This modular structure enables users to easily and transparently adjust assumptions throughout the elicitation process. This includes modifying expert input, altering loss components, or incorporating post-hoc evaluation procedures. This possibility supports a systematic exploration of the elicitation problem under varying assumptions and contexts. Consequently, \pkg{elicito} may also be of interest from a methodological research perspective, as it enables the study of relationships between different input–output structures for a given problem.

Accordingly, the target audience of \pkg{elicito} includes both researchers developing prior elicitation methodologies and applied practitioners aiming to construct prior distributions for specific use cases. While the current implementation of \pkg{elicito} remains relatively technical, future development will focus on creating higher-level interfaces (i.e., facades) tailored to common practical applications. These interfaces aim to simplify user interaction with the core \code{Elicit} object by incorporating default assumptions at appropriate stages of the workflow. Although this design choice may reduce transparency to some extent, it is expected to improve usability and enhance accessibility for end users \citep{simpson2017penalising}. In this context, future work should also include the development of user guidelines for specific sub-tasks or problem types encountered in the prior elicitation process. One example is the handling of non-uniqueness which arises primarily from limitations in expert information, either in quantity or quality \citep{lopez2015nonlinear}. Several strategies can be employed to address this challenge. These include modifying the loss function, for example, by incorporating additional expert input or introducing structural priors via regularization, reparameterizing the statistical model to reduce the dimensionality of the parameter space or to increase the influence of the parameters on the outcome variable, and applying post-hoc methods that aggregate across the solution space to derive a representative solution that appropriately reflects the underlying uncertainty. This knowledge about solution strategies should be made available in form of user-guidelines.

A central challenge in the development of prior elicitation methods lies in balancing the use of default assumptions with the incorporation of individualized expert feedback. On one hand, elicitation methods should minimize the cognitive, temporal, and financial demands placed on experts \citep{mikkola2023prior}. On the other hand, maintaining a ``human-in-the-loop'' is essential to ensure that the resulting prior distributions align with the expert’s expectations \citep{hartmann2020flexible}. Expressing prior beliefs in quantitative terms is inherently non-trivial. An iterative feedback mechanism is therefore essential, enabling users to refine their input through a process of trial and error. While \pkg{elicito} represents the key steps of the prior elicitation workflow as modular, algorithmic components, it is not intended to fully automate the elicitation process. Instead, its primary objective is to make explicit the input information required to construct prior distributions,  thereby facilitating a systematic and transparent approach to this inherently complex task.

Finally, \pkg{elicito} offers substantial scope for future enhancements and continued development. Two key directions for extension are particularly noteworthy. First, expanding compatibility with multiple probabilistic programming languages is a central objective. At present, user-defined functions must rely on the TensorFlow backend to ensure integration within the computational graph and enable correct backpropagation. However, the conceptual foundation of our simulation-based framework should ideally remain independent of any specific implementation, so as not to constrain the potential user base or limit interoperability. The second major extension involves transforming the current workflow into a fully Bayesian framework, thereby enabling the inference of posterior distributions over hyperparameters. This would enable a more accurate representation of uncertainties introduced throughout the elicitation process, from initial knowledge extraction to the fitting of prior distributions.

\newpage
\section*{Declarations}
\paragraph{Acknowledgments}
We thank Luna Fazio for the long discussions with valuable comments and suggestions that greatly improved this work.
\paragraph{Funding}
Not applicable.
\paragraph{Competing interests}
The authors have no competing interests to declare that are relevant to the content of this article.
\paragraph{Consent to participate}
Not applicable.
\paragraph{Consent for publication}
Not applicable.
\paragraph{Data \& Materials availability}
The simulation results and notebooks with an implementation of the case study are available in the supplementary material on GitHub \url{https://github.com/florence-bockting/elicito-software-paper}.
\paragraph{Code availability}
Code underlying the case stduy is based on/provided by our Python package \emph{elicito} \citep[][v0.6.0]{bockting2025elicito}. To ensure the reproducibility of our results, the specific version of \emph{elicito} used in this study has been archived on Zenodo and is accessible via the following DOI \url{https://doi.org/10.5281/zenodo.15671710}.
\paragraph{Author contribution}
Conceptualization: FB, Methodology: FB, Software: FB, Writing (Original Draft): FB, Writing (Editing): FB, Writing (Review): FB, Visualization: FB, Supervision: PB

\newpage
\bibliography{references}

\newpage
\section*{Appendix}
\appendix
\section{Symbols and their definition}
\label{app:symbols}
\begin{table}[ht]
    \centering
    \begin{tabular}{p{0.07\textwidth} p{0.32\textwidth} p{0.4\textwidth}}
    \hline
        \emph{Symbol} & \emph{Description} & \emph{Comment}  \\
    \hline
         $\theta_p$ & model parameter & $p=1,\ldots,P$\\
         $\lambda_k$ & model hyperparameter & $k=1,\ldots,K$ \\
         $p_\lambda(\theta)$ & priors parameterized by $\lambda$ & in case of non-parameteric priors $\lambda$ refer to the parameters of the deep generative model\\
         $\mathcal M(\lambda)$ & generative model incl. priors & for short $\mathcal M$\\
         $\mathcal{P}_{pr}$ & prior specification & $pr = (\text{parametric}, \text{non-parametric})$ \\
         $z_i$ & target quantity & defined as $z_i=c_i(\lambda)$ with $i=1,\ldots,I$ \\
         $t_m$ & simulated elicited statistics & defined as $t_m=f_j(z_i)$ with $m=(i,j)$\\
         $f_j$ & elicitation technique & $j=1,\ldots,J$\\
         $\hat t_m$ & expert-elicited statistic & \\
         $\mathcal{D}_m$ & discrepancy measure & \\
         $L_m$ & loss component & $L_m=w_mD_m(t_m(\lambda),\hat t_m)$ \\
         $w_m$ & weight of loss component & \\
         $\mathcal{L}(\lambda)$ & total loss wrt $\lambda$ & $\mathcal{L}(\lambda)=\sum_{m=1}^ML_m$\\
         $E$ & epochs & \\
         $B$ & batch size & default $B=128$ \\
         $M$ & number of prior samples & default $M=200$\\
    \hline
    \end{tabular}
    \caption{List of symbols and their definitions used throughout the paper}
    \label{tab:symbols}
\end{table}
\newpage

\section{Implementation of parametric-prior example}\label{app: impl-param-model}
\subsection{Implementation of design matrix}
\label{app: design-matrix}
\begin{CodeChunk}
    \begin{CodeInput}   
    # design-matrix for three-level dummy-coded predictor
    def design_categorical(n_gr):
        incpt = [1]*3
        contrast_01 = [0, 1, 0]
        contrast_02 = [0, 0, 1]
         
        # contrast matrix
        c = tf.stack([incpt, contrast_01, contrast_02], axis=-1)
         
        # design matrix
        x = tf.concat([tf.broadcast_to(c[i, :], (n_gr, c.shape[1])) for 
                      i in range(c.shape[0])], axis=0)
        return tf.cast(x, tf.float32)
    \end{CodeInput}
    
    \begin{CodeOutput}
    # example output of design matrix for n_gr=1
    > design_categorical(n_gr=1)
    
    <tf.Tensor: shape=(3, 3), dtype=float32, numpy=
    array([[1., 0., 0.],
           [1., 1., 0.],
           [1., 0., 1.]], dtype=float32)>
    \end{CodeOutput}
\end{CodeChunk}

\newpage
\section{Additional results of toy example 2}
\label{app: results-toy-example-2}
\begin{figure}[h]
        \centering
        \includegraphics[width=0.8\linewidth]{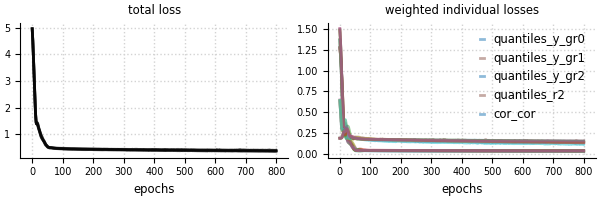}
        \caption{Convergence of the total loss (left) and the single loss components (right). Plot is created using \code{el.plots.loss(eliobj\_deep)}.}
        \label{fig:loss-deep-prior}
    \end{figure}
\begin{figure}[h]
        \centering
        \includegraphics[width=0.8\linewidth]{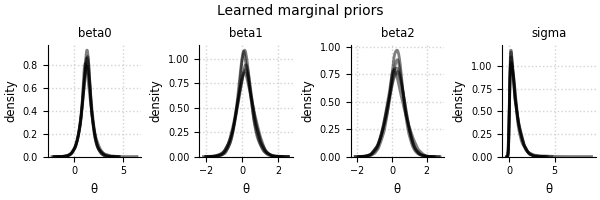}
        \caption{Learned marginal prior distribution for all parallel run. The plot is generated using \code{el.plots.prior\_marginals(eliobj\_deep)}.}
        \label{fig:prior-marginals-deep}
    \end{figure}
\begin{figure}[ht]
        \centering
        \includegraphics[width=0.8\linewidth]{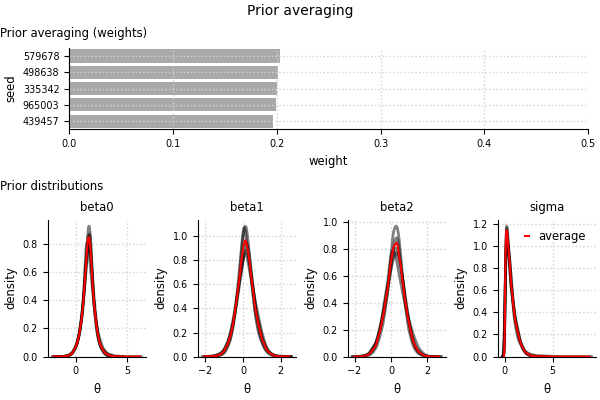}
        \caption{Prior averaging across multiple parallel runs. Upper panel demonstrates weights associated with each replication used for computing the prior average. Lower panel shows the individual marginal priors as well as the averaged prior (in red). Plot is created using \code{el.plots.prior\_averaging(eliobj\_deep)}.}
        \label{fig:prior-averaging-deep}
    \end{figure}
    
\end{document}